\documentclass[prl,twocolumn,showpacs,floatfix,superscriptaddress,nofootinbib]{revtex4-1}
\pdfoutput=1
\usepackage{graphicx}
\usepackage{times}
\usepackage{amsmath,amssymb,bm}
\usepackage{dcolumn}
\usepackage{color}

\begin{document}

\title{A Panoply of Orders from a Quantum Lifshitz Field Theory}
\author{Leon Balents}
\affiliation{Kavli Institute of Theoretical Physics, University of California, Santa Barbara, Santa Barbara, CA 93106}
\author{Oleg A. Starykh}
\affiliation{Department of Physics and Astronomy, University of Utah, Salt Lake
City, UT 84112}
\date{October 25, 2015}
\begin{abstract}
  We propose a universal non-linear sigma model field theory for one
  dimensional frustrated ferromagnets, which
  applies in the vicinity of a ``quantum Lifshitz point'', at which
  the ferromagnetic state develops a spin wave instability.  We
  investigate the phase diagram resulting from perturbations of the
  exchange and of magnetic field away from the Lifshitz point, and
  uncover a rich structure with two distinct regimes of different
  properties, depending upon the value of a marginal, dimensionless,
  parameter of the theory.  In the regime relevant for one dimensional
  systems with low spin, we find a metamagnetic transition line to a
  vector chiral phase.  This line terminates in a critical endpoint
  from which emerges a cascade of multipolar phases.  We show that the
  field theory has the property of ``asymptotic solubility'', so that
  a particular saddle point approximation becomes asymptotically exact
  near the Lifshitz point.  Our results provide an analytic framework
  for prior numerical results on frustrated ferromagnets,
  and can be applied much more broadly.
\end{abstract}
\maketitle

The study of order in all its variety anchors the field of condensed
matter physics.  Some current goals at the vanguard of this enterprise
include characterizing ``hidden'' orders, determining the mechanism
behind ``competing'' or ``intertwined'' orders, and understanding
quantum phase transitions between different orders.  These problems
arise in diverse systems ranging from frustrated quantum magnets to
correlated electron materials like the cuprates.  Various notions of
topological order, entanglement, and relativistic/conformal field theory
have been perhaps the most common avenues to investigate
these phenomena.  Yet the issues persist and gain relevance from the
accumulation of experiments, and for the most part resist the attack
by these approaches.

Here we describe a different route which unifies the three above
themes in a tangible context within quantum magnetism.  Specifically,
we study a {\em quantum Lifshitz} transition between a ferromagnet and
a spiral magnet or quantum paramagnet, which is realized for example
in the well-studied Frustrated Ferromagnetic Heisenberg Chain (FFHC):
\begin{equation}
  \label{eq:3}
  H_{FFHC} = \sum_n \left[ - {\bm S}_n \cdot {\bm S}_{n+1} + \beta
    {\bm S}_n \cdot {\bm S}_{n+2} - h S_n^z \right].
\end{equation}
With increasing frustration $\beta$, Eq.~\eqref{eq:3} has a Lifshitz
point at $\beta=1/4$, $h=0$.  
Numerical studies of the FFHC have previously demonstrated that
metamagnetism and a rich sequence of {\em multipolar} phases -- a type
of hidden order which does not appear in spin-spin correlation
functions -- appear in the vicinity of this point for non-zero applied
magnetic field $h$.  The simplest of these
phases is the (spin) angular momentum $p=2$ multipole, or quadrupolar state, also known as a spin nematic, which
breaks spin rotational symmetry but preserves invariance with respect
to time reversal \cite{Andreev1984}.  As such, the spin nematic is
characterized by an order parameter bilinear in the microscopic spins.
It can be understood as a state of bound, condensed pairs of
magnons\cite{Chubukov1991,Shannon2006,Kecke2007,Hikihara2008,Sudan2009,mzh2010}.
The spin nematic has been sought experimentally in a number of
quasi-one-dimensional materials which approximately realize the FFHC
\cite{Svistov2011,Masuda2011,mourigal2012,Nawa2013,Nawa2014,Prozorova2015,drechsler2015}.

Theoretically, the proliferation of multipolar phases with $p>2$ near
the Lifshitz point in the FFHC is most extraordinary, and begs
theoretical explanation.  We provide it by formulating a
non-relativistic Non-Linear Sigma Model (NLSM) with dynamic critical
exponent $z=4$, which governs this transition.  
An asymptotically exact analytic solution of the Lifshitz NLSM
produces the line of the first-order metamagnetic transitions which terminate 
at the metamagnetic end-point beyond which the multi-particle condensation transition turns continuous.
This condensation produces a remarkable cascade
of multipolar states with very large multipoles. 
From a formal perspective the cascade is quite unusual: a single field theory describes a
collection of phases whose order parameters are formed from {\em arbitrarily} large powers of the fundamental 
(spin) fields of the theory.  

{\em Lifshitz non-linear sigma model:}  Instead of focusing on a
specific microscopic model such as the FFHC in Eq.~\eqref{eq:3}, we
introduce a universal quantum field theory description which is based
on translational symmetry and SU(2) spin-rotation invariance.  Since
we are interested in continuous transitions out of a ferromagnet,
whose magnetization is O(1) and quantized given SU(2) symmetry, we
expect that locally there is a (possibly fluctuating) magnetization,
even close to and on both sides of the
quantum critical point.  
Hence we propose that the low-energy properties of
the system are described by a non-linear sigma model (NLsM) formulated in terms of unit vector $\hat{m} = (\hat{m}_1, \hat{m}_2, \hat{m}_3)$ which describes 
magnetization density. The action is 
\begin{eqnarray}
  \label{eq:1}
  S & = & \int \! dx d\tau \big\{ is{\mathcal A}_B[\hat{m}] - \delta
    |\partial_x \hat{m}|^2 + \kappa |\partial_x^2 \hat{m}|^2 \nonumber \\
    && + \upsilon |\partial_x \hat{m}|^4 - h \hat{m}_z\big\}.
\end{eqnarray}
Here $s$ is the spin and ${\mathcal A}_B$ is the Berry phase term
describing those spins (implicitly a factor of the inverse lattice
spacing, set to unity, is present in this coefficient, which
compensates for the dimension of length due to the $x$ integral).  It can be
written in various ways, for example \cite{Schlittgen2001},
\begin{equation}
  \label{eq:2}
  {\mathcal A}_B = \int_0^1 du\,  \hat{m}\cdot \partial_\tau \hat{m}  \times \partial_u \hat{m} 
  = \frac{\hat{m}_1 \partial_\tau \hat{m}_2 - \hat{m}_2 \partial_\tau \hat{m}_1}{1+ \hat{m}_3}, 
\end{equation}
where we introduced a fictitious auxiliary coordinate $u$ such that
$\hat{m}(u=0) = \hat{z}$ and $\hat{m}(u=1) = \hat{m}$ is the physical
value. The main important point is that ${\mathcal A}_B$ contains a
single derivative with respect to imaginary time $\tau$.  

The action $S$ contains all leading terms in gradients of $\hat{m}$.
The parameter $\delta$ ($\propto \beta-1/4$ in the FFHC) tunes the zero
field criticality: a trivial fully ordered ferromagnetic (FM) state with
constant $\hat{m}$ and no fluctuations obtains for $\delta<0$, while
the system is non-trivial for $\delta>0$.  The absence of fluctuations
for $\delta<0$ is due to the $\mathcal{A}_B$ term, which makes the
dynamics completely different from the commonly studied relativistic
NLsM's.  Further, note that there are two terms, $\kappa$ and $\upsilon$, quartic in
derivatives, which is crucial in the following.  The $\upsilon$ term has been
ignored in previous field theoretic
approaches\cite{Kolezhuk2002,Sirker2011}.

The action \eqref{eq:1} needs a condition for stability against large
gradients of $\hat{m}$.  Starting from constraint $\hat{m}\cdot \hat{m}=1$,
it is easy to obtain 
$|\partial_x^2 \hat{m}|^2 > |\partial_x \hat{m}|^4$, which is enough
to show stability is present so long as $\upsilon+\kappa>0$.  This means 
{\em negative} $\upsilon$ in \eqref{eq:1} is allowed so long as $\upsilon > -\kappa$.  

The action describes several distinct dynamical regimes.  For
$\delta<0$, the excitations above the ground states are quadratically
dispersing spin waves, $\omega \sim k^z$, characterized by the dynamical critical
exponent $z=2$, which is easily seen by equating the linear $\tau$
derivative in $\mathcal{A}_B$ with the second spatial derivative in
the $\delta $ term.  For $\delta=0$, the dynamics changes to $z=4$.
For $\delta>0$, the theory is more non-trivial, and there is even a
$z=1$ regime (see below).  

{\em Asymptotic solubility:} Physically, the absence of fluctuations
in the FM state suggests a saddle point approximation may apply near
to it.  Indeed, a simple rescaling $x \rightarrow \sqrt{\kappa/\delta} ~x'$
and $\tau \rightarrow \kappa \tau'/\delta^2$ transforms the action into
suggestive form (we defined $v=-\upsilon/\kappa$ and $h' = h\kappa/\delta^2$)
\begin{eqnarray}
  \label{eq:4}
  S & = & \sqrt{\frac{\kappa}{\delta}} \int \! dx' d\tau'\, \big\{
          is{\mathcal A}'_B[\hat{m}] - {\rm sign}(\delta)|\partial_{x'} \hat{m}|^2 + |\partial_{x'}^2 \hat{m}|^2 \nonumber \\ && -
  v|\partial_{x'} \hat{m}|^4 - h' \hat{m}_z\big\} ,
\end{eqnarray}
which shows that near the critical point, when $\delta/\kappa \ll 1$, the
action is large in dimensionless terms so that a saddle point analysis
becomes asymptotically correct on approaching the Lifshitz point (the
prime on the Berry phase term simply indicates that it includes the
time derivatives inside are taken with respect to $\tau'$).  Because
$|\delta|$ appears only in the prefactor of the action in
Eq.~\eqref{eq:4}, the phase diagram at the saddle point level and only
the dimensionless parameters $v$ and $h'$ control the saddle point.
Note that $v<1$ defines the stability region of the theory.

\begin{figure}[h!]
   \centering
    \includegraphics[width=0.35\textwidth]{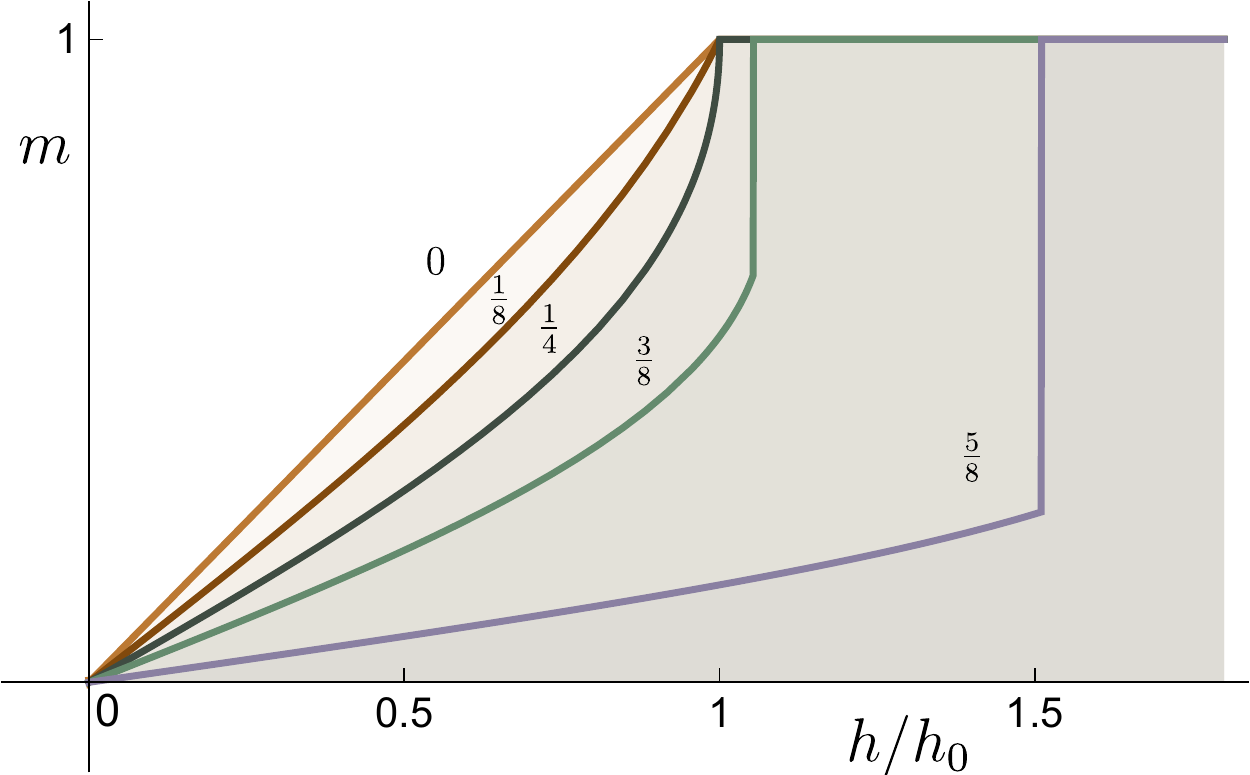}
     \caption{Saddle point result for the magnetization $m(h)$ for different values of interaction
     parameter $v$, which is shown next to each curve. }
     \label{fig:m-h}
\end{figure}

The saddle point of Eq.~\eqref{eq:1} with minimum action describes a cone
(umbrella) state:
\begin{equation}
  \label{eq:5}
  \hat{m}_{\rm sp} = (\varphi\cos q x, \varphi\sin qx,
\sqrt{1-\varphi^2}),
\end{equation}
with $0\leq \varphi\leq 1$ and $q$ functions of the parameters of the
action.  Solutions with both sign of $q$ are degenerate, which
reflects spontaneous breaking of reflection symmetry and chiral order:
$\hat{z}\cdot \hat{m}_{\rm sp}\times \partial_x\hat{m}_{\rm sp} =
\varphi^2 q \neq 0$.
For sufficient large field, $h>h_c$, the solution is simply the
ferromagnetic one, with $\varphi=0$.  On reducing the field, there are
two possible behaviors.  For $\upsilon>-\kappa/4$ ($v<1/4$), a continuous
transition occurs at the critical field $h_c=h_0= \delta^2/(2\kappa)$.  The
``order parameter'' $\varphi$, which represents the local moment
transverse to the magnetic field, increases smoothly from zero below
$h_0$.  This corresponds to the point of local instability of the FM
phase to single magnons, which Bose condense when their energy
vanishes at $h_0$.  For $\upsilon<-\kappa/4$ ($v>1/4$), the transition occurs
discontinuously at $h_c>h_0$, at which point the ferromagnetic state
is still locally stable.  The order parameter jumps to a non-zero
value $\varphi_c$ for $h=h_c-0^+$.  This is a metamagnetic transition,
described by
\begin{equation}
  \label{eq:9}
  \varphi_c^2 = \frac{2\sqrt{v}-1}{v}, h_c = \frac{\delta^2}{8 \kappa \sqrt{v}(1-\sqrt{v})}, q_c^2 = \frac{\delta}{4\kappa(1-\sqrt{v})},
\end{equation}
which hold for $1/4<v<1$.   Due to the aforementioned scale
invariance, the metamagnetic line extends for all $\delta$ at the
saddle point level.   We emphasize that the saddle point results are
asymptotically exact, and hence provide direct predictions for
experiment for systems close to the Lifshitz point.  For example, the saddle-point behavior of the magnetization $m=\sqrt{1-\varphi^2}$ is shown in Fig.~\ref{fig:m-h}.

{\em Quantum corrections:} Fluctuations beyond the saddle point 
have several types of effects.  One innocuous effect is that of
phase fluctuations within the ``cone phase'': configurations of form
of Eq.~\eqref{eq:5} with $qx \rightarrow qx + \theta$ have
small action when $\theta(x,\tau)$ has small space-time gradients.
Fluctuations of $\theta$ are thereby described by a free boson theory
with central charge $c=1$, which converts the long-range cone order
into power-law spin correlations, but preserves the chiral order.
These properties characterize a ``vector chiral'' phase (VC), 
identified previously in the FFHC.

A more drastic effect of fluctuations is to move the phase boundaries
and even introduce new phases.  This is due to the differing
contribution of fluctuations to the ground state energy of different
states.   Quantum fluctuations modify the energy of the cone state but
do not affect that of the (fluctuationless) ferromagnetic state.
Hence fluctuations may shift the FM-cone
transition to lower magnetic field.   Remembering that $h_0$ is the
single magnon condensation field, it makes sense to ask if quantum
fluctuations can lower $h_c$ all way down to $h_0$? Note that
affirmative answer to this question implies the appearance of the
metamagnetic {\em endpoint} beyond which the FM-cone transition
becomes continuous.   

To investigate this question, we write the magnetization $\hat{m}$ in the {\em co-moving}
system of coordinates 
\begin{equation}
\hat{m} = \sqrt{2 - \frac{\bar{\eta}\eta}{s}} [\frac{\bar{\eta} + \eta}{2\sqrt{s}} \hat{e}_1 + i \frac{\bar{\eta} - \eta}{2\sqrt{s}} \hat{e}_2] + 
(1 - \frac{\bar{\eta}\eta}{s}) \hat{e}_3 ,
\label{eq:rotating-m}
\end{equation}
where the rotating dreibein $\hat{e}_j(x)$ are chosen as follows:
$\hat{e}_1 \times \hat{e}_2 = \hat{e}_3 \equiv \hat{m}_{\rm sp}$.
The fields $\bar{\eta}, \eta$ describe magnons, transverse fluctuations of the
magnetization.  
To quadratic order  the action
in Eq.~\eqref{eq:1} becomes $S = \int\! d\tau \left[ \int dx\, 
\bar{\eta} \partial_\tau \eta + H_{\rm fluct}\right]$, which
shows that $\bar{\eta},\eta$ are canonical Bose operators, and
$H_{\rm fluc}(\bar{\eta},\eta)$ is a Hamiltonian. Fourier transforming it into momentum
space shows that $H_{\rm fluc}$ contains both normal and anomalous
terms:
\begin{equation}
H_{\rm fluc} = \sum_k 2A_k \bar{\eta}_k \eta_k + B_k (\eta_k \eta_{-k} + \bar{\eta}_k \bar{\eta}_{-k}) .
\label{eq:H_fluc}
\end{equation}
Here coefficients $A_k, B_k$ are functions of momentum $k$ and depend on parameters $\delta, \kappa, v, h$ and $\varphi$
of the saddle point action. Diagonalization of \eqref{eq:H_fluc} with
the help of a standard Bogoluiubov transformation
gives us the desired correction: the zero-point energy
$\delta {\cal E}_{\rm cone} = \frac{1}{N} \sum_k \{ \sqrt{A_k^2 - B_k^2} - A_k\}$.

We use this corrected energy to identify a metamagnetic endpoint.  A
metamagnetic endpoint occurs at $\delta=\delta_c$ if, for
$\delta>\delta_c$, the cone state remains higher in energy than the FM
state for all $h\geq h_0$, while for $\delta<\delta_c$, the cone state
has lower energy than the FM one for some range of fields $h_0<h<h_c$.
Hence the endpoint is determined by the condition that the energy of
the cone state equals that of the FM state at $h=h_0$, i.e.
$\Delta E = \Delta {\cal E} -\delta {\cal E}_{\rm cone}=0$ at $h=h_0$
where the first term
$\Delta {\cal E} = {\cal E}_{FM}- {\cal E}_{\rm cone}$ represents the
saddle point energy difference, and the last is the Bogoliubov
correction.

Before analyzing this in detail, we note that from Eq.~\eqref{eq:4}, the
fluctuation corrections to the energy are expected to be reduced from the saddle
point value by a factor of $\sqrt{\delta/\kappa}$, which is assumed small
for consistency of the approach.  Hence they can affect
the balance between cone and FM states only when the energy difference
between the two is already small at the saddle point level.  Therefore we
now focus on the regime close to the onset of metamagnetism, and let
$v=1/4+\epsilon$ in what follows, with $\epsilon \ll 1$.  In this
limit, $\Delta {\cal E}(h_0) = \frac{256}{27}\kappa \epsilon^3 (\delta/\kappa)^2$.

The fluctuation correction $\delta {\cal E}_{\rm cone}$ contains a
regular cutoff-dependent part and a singular universal term.  The
former may be absorbed into a renormalized coupling
$v \rightarrow \tilde{v}$ and likewise $\epsilon$.  The latter
represents a physically distinct contribution to the cone state
energy.  For the lattice FFHC it was obtained previously in
\cite{Krivnov1996}.  We obtain
$\delta {\cal E}_{\rm cone}^{\rm sing} = s^{-1} \int_{-\infty}^\infty \frac{d k}{2\pi} \frac{100 \delta^3 \epsilon^2}{\kappa^2 k^2 + 2 \kappa \delta} = 
(25\sqrt{2}/s) \kappa \epsilon^2 (\delta/\kappa)^{5/2}$.

Now combining the saddle point and corrections, we find that the total energy
$\Delta E = \Delta {\cal E}(h_0) - \delta {\cal E}_{\rm cone}^{\rm sing}$
is seen to change sign at $\delta_c \approx 0.07 \kappa s^2 \epsilon^2$,
indeed indicating a metamagnetic endpoint. Since $\delta_c \ll 1$ with
$\epsilon\ll 1$, this is within the regime of validity of the field
theory.  

\begin{figure}[h!]
  \centering
    \includegraphics[width=0.45\textwidth]{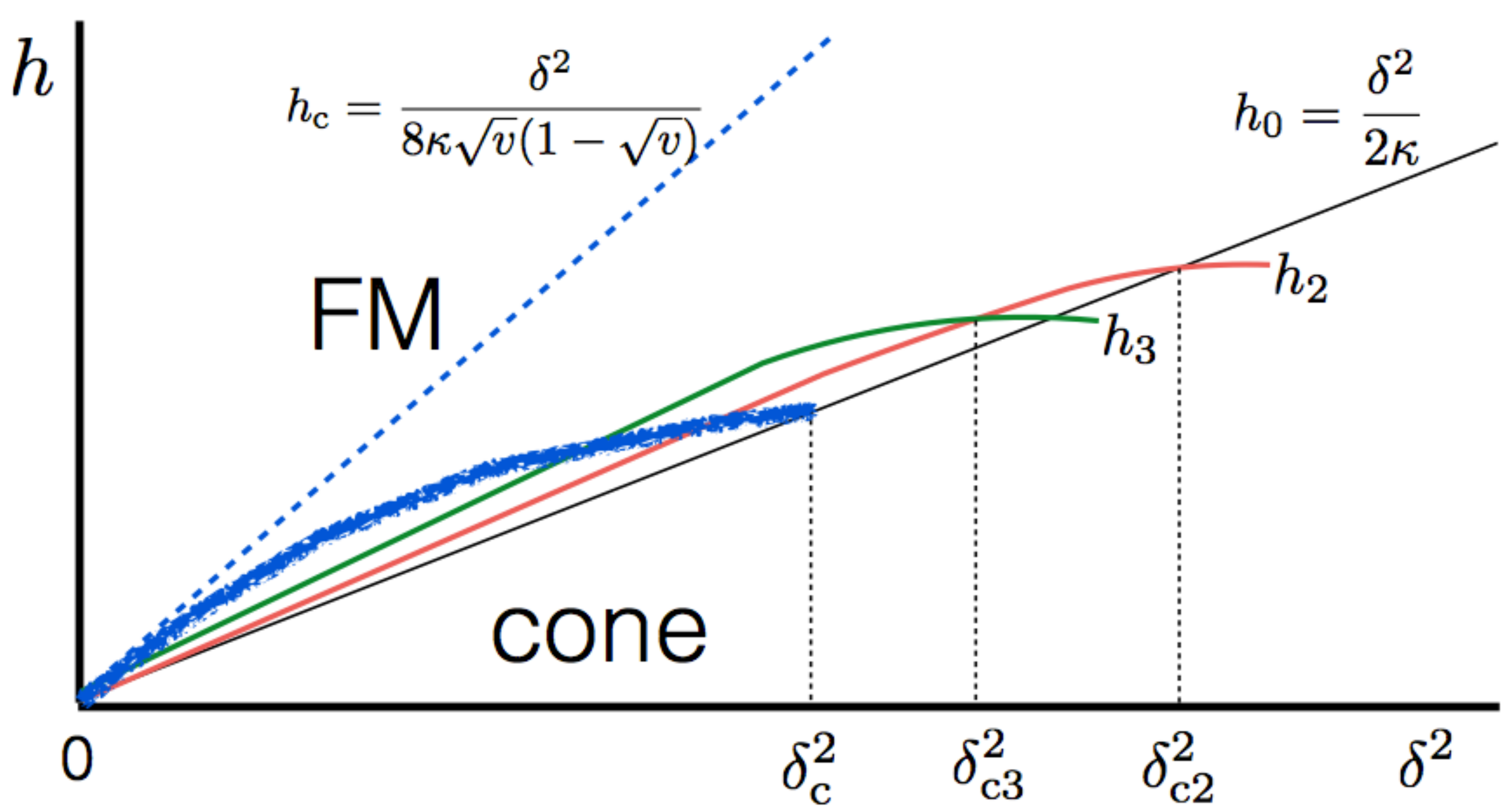}
    \caption{Stability curves (schematic).  The thin dashed (blue)
      line shows the critical $h_{\rm c}$ field of the first order
      transition within the classical saddle point approximation.  The
      wide brushed (blue) line indicates $h_{\rm c}$ as modified by
      quantum fluctuations. It crosses the thin (black) single-magnon
      instability field $h_0$ at $\delta=\delta_c$. The red (green)
      solid lines denote the critical magnetic fields $h_2 (h_3)$
      describing two- (three-) magnon condensation instabilities. The
      $h_3(\delta)$ curve is a conjecture, see text for more details.}
     \label{fig:phase}
\end{figure}

{\em Quantum few-body physics:} Considering the above result, we see
that for $\delta>\delta_c$, the effective attraction between magnons
is too weak to induce collapse.  Hence one might conclude that the
first instability of the ferromagnet upon reducing the field $h$ is to
continuous single-magnon condensation at $h=h_0$.  Here we argue this
is incorrect, because there is a third possibility.  While the
attraction for $\delta>\delta_c$ is too weak to induce collapse, it
still is strong enough to produce bound states of a finite number of
magnons, which leads to distinct multipolar phases in a range
$\delta_c <\delta< \delta_{c2}$, that set in at $h>h_0$.

As we consider larger $\delta$, the semiclassical analysis becomes
inadequate, and a full quantum treatment of the action in
Eq.~\eqref{eq:1} becomes necessary, which is daunting due
to its non-polynomial nature (implicit in the NLsM constraint).
In principle, by using Eq.~\eqref{eq:rotating-m} with $\hat{e}_\mu =
\hat{x}_\mu$, one can expand and truncate the action to
$\mathcal{O}(\eta^{2n})$ for an exact treatment of $n$-magnon states,
since higher order terms, if properly normal-ordered, annihilate these
states.  
This leads to a quantum Hamiltonian for bosonic fields
$\overline{\eta},\eta$ with an unconventional kinetic
energy and up to $n$-body momentum dependent interactions.   Due to
the complexity of this problem, we have limited ourselves to the $n=2$
case.  This expansion yields
\begin{eqnarray}
  \label{eq:6}
  H & = &  \sum_k \epsilon_k \overline{\eta}_k \eta^{}_k
         \\
  && +
  \frac{1}{2L} \sum_{kpp'} V(k,p,p')\overline{\eta}_{k/2+p}
  \overline{\eta}_{k/2-p} \eta^{}_{k/2-p'}\eta^{}_{k/2+p'}, \nonumber
\end{eqnarray}
with $\epsilon_k = (h + 2\kappa k^4 - 2\delta k^2)/s$ and $V(k,p,p')$  in \cite{suppl}.

One can gain some insight by focusing on the minima of $\epsilon_k$,
which occur at $k=\pm q$, with $q=\sqrt{\delta/(2\kappa)}$.
We therefore define new fields $\psi_{a,k} = \eta_{(2a-3)q + k}$ for
$|k| \ll q$ and $a=1,2$.  Then, Fourier transforming back to real
space, one obtains, {\em assuming all the scattered magnons remain
  near the two minima},
\begin{eqnarray}
  \label{eq:14}
  H & = & \int dx \, \Big\{\sum_{a=1}^2 \overline{\psi}_a 
    (\epsilon_0 - \frac{\partial_x^2}{2m} )\psi^{}_a +\\
  & & + \frac{1}{2}\gamma_1 \left[ (\overline{\psi}_1 \psi^{}_1)^2 + (\overline{\psi}_2
    \psi^{}_2)^2\right] + \gamma_2 \overline{\psi}_1 \psi^{}_1
  \overline{\psi}_2\psi^{}_2 \Big\}, \nonumber
\end{eqnarray}
where $\epsilon_0 = h/s- \delta^2/(2\kappa s)$,
$m=s/8\delta$, $\qquad \gamma_2 = \delta^2 (5-4v)/(\kappa s^2)$ and $\gamma_1 = \delta^2 (1-4v)/(2\kappa s^2)$.
Observe that for $v>1/4$, when the saddle point analysis found
metamagnetism, the {\em intra}-valley interaction
$\gamma_1$ is negative, i.e. {\em attractive}.  
As is well known, bosons with attractive delta-function potential, such as described by the $\gamma_1$ term in \eqref{eq:14},
undergo collapse \cite{Calogero1975,Kosevich1990,Krivnov1996} -- the ground state of the system is given by the $N$-body bound state
in which all $N$ bosons of the system participate.  This collapse
corresponds to the metamagnetic transition.  In reality an infinite
collapse is prevented by three-body interactions, and moreover the
saddle point condition is renormalized with increasing $\delta$ as we
found above, leading to the metamagnetic endpoint.

We can investigate renormalizations at the
two-body level from Eq.~\eqref{eq:6}.   In particular, taking the {\em full} dispersion and
momentum-dependent interactions, we solve the two-body Schr\"odinger
equation for the minimum energy state.  The general form for such a
state is $|\psi,k\rangle = \int \! \frac{dq}{2\pi}
\Psi(q;k)\overline{\eta}_{k/2+q}\overline{\eta}_{k/2-q} |0\rangle$,
where $|0\rangle$ is the boson vacuum, i.e. the ferromagnetic state,
$k$ is the (conserved) center of mass momentum, and the two-magnon
wavefunction obeys
\begin{eqnarray}
  \label{eq:7}
&&  (\epsilon_{k/2+p} + \epsilon_{k/2-p} - E) \Psi(p;k) + \int \!
  \frac{dp'}{2\pi} V(k,p,p') \Psi(p';k)\nonumber\\
  &&=0.
\end{eqnarray}
This equation can be solved exactly \cite{suppl}.  We obtain the minimum energy
state for $k=\pm 2q$, which corresponds to a pair of magnons from the
same minima, and find the binding energy $\epsilon_b=2\epsilon_q - E$
given by the relation
\begin{equation}
  \label{eq:8}
  \sqrt{\epsilon_b} \approx \sqrt{\epsilon_{b0}} \left[1 -
   \left(\frac{\delta}{\delta_{c2}}\right)^{1/2}\right]+ \mathcal{O}(\delta^{5/2}),
\end{equation}
where $\epsilon_{b0} = \epsilon^2 \delta^3/(8\kappa^2 s^3)$ is just the
na\"ive binding energy one would obtain from the delta-function
interaction model, $\epsilon_{b0} = m\gamma_1^2/4$, and the term in
the brackets represents the leading correction.  This defines a
critical value $\delta_{c2} = \frac{128}{625}\kappa s^2 \epsilon^2 \approx 0.2
\kappa s^2 \epsilon^2$, such that the two-magnon bound state
disappears for $\delta>\delta_{c2}$.   

\begin{figure}[h!]
    \centering
    \includegraphics[width=0.42\textwidth]{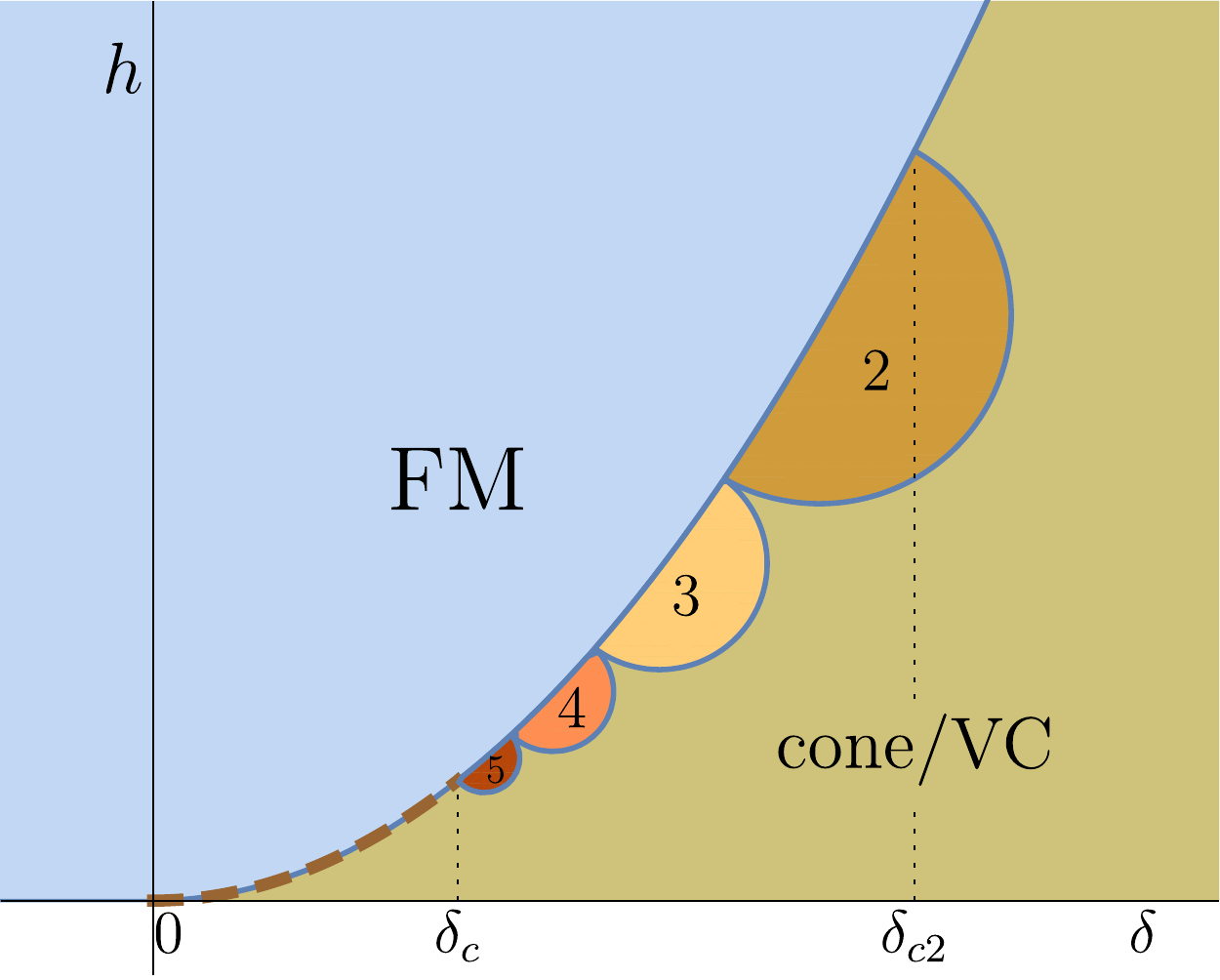}
     \caption{Schematic phase diagram in $\delta-h$ plane. The dashed
       line is the metamagnetic transition emerging from the Lifshitz
       point at the origin.  ``FM'' and ``cone/VC''  denote the fully
       polarized ferromagnetic and cone/vector chiral phases, respectively.  Integers
       $2,3,4,5$ label multipolar phases comprised of the
       corresponding number of bound magnons.   }
     \label{fig:multipolar}
\end{figure}

Importantly, we note that $\delta_{c2}>\delta_{c}$, which implies that
in this interval the ferromagnetic state is unstable to two-magnon
condensation for a non-zero range of fields $h>h_0$.  In principle, we
should now check for bound states of more than two magnons.
Unfortunately, we have not been technically able to accomplish this.
We speculate that in the range $\delta_c<\delta<\delta_{c2}$,
bound states of increasing numbers of magnons appear with decreasing
$\delta$, at thresholds $\delta_{c,n}$, with
$\delta_c<\delta_{c,n}<\delta_{c,n'}$ for $n>n'$.  
\footnote{Note that strictly speaking the actual metamagnetic endpoint $\delta_{\rm c}^{*}$ is determined by the crossing of the
renormalized first-order transition field $h_{\rm c}$ with $h_{\rm n_{\rm max}}$, the field 
of the maximum-possible $n_{\rm max}$-complex condensation. Fig.~\ref{fig:phase} shows
that $\delta_{\rm c}$ provides an upper bound on $\delta_{\rm c}^{*}$.}
This would imply a
sequence of distinct multipolar phases just below saturation in this
intermediate range of $\delta$, as shown schematically in
Fig.~\ref{fig:multipolar}. Note that the defining feature of the $n^{\rm
  th}$ multipolar phase is the presence of a gap for excitations with
spin $S^z <n$.  In one dimension, due to fluctuations, there is no
true multipolar condensate, and each phase evolves smoothly from more
condensate-like to spin-density-wave-like on reducing
field \cite{Hikihara2008,Starykh2014}.   

{\em Microscopic calculation of $v$:} The crucial dimensionless
parameter $v$ of the theory cannot be determined within our field
theory approach.  We found two ways to fix its value by comparing
field theory predictions with those of complimentary microscopic
calculations \cite{suppl}.  In the first, large spin $s \gg 1$
calculation, we use the standard spin-wave technique to calculate the
leading spin-wave corrections to the ground state energy and the
optimal spiral wave vector of the spin-$s$ $J_1-J_2$ chain.  Comparing
these results with the saddle point analysis, we find $v = 3/(2s)$.
Hence $v < 1/4$ for large $s$, and thus metamagnetism occurs only for
spin chains with $s < s_c = 6$, in agreement with earlier
Bethe-Salpeter calculations \cite{Arlego2011,Kolezhuk2012}.

For the $s=1/2$ chain, we match the value of the order parameter jump
$\varphi_c$, \eqref{eq:9}, at the metamagnetic transition to the
corresponding value of the magnetization $m_c = (\sqrt{7}-1)/3$
reported in Ref.~\onlinecite{Krivnov1996}. This gives, via
$m_c^2 = 1 - \varphi_c^2$, that $v_{s=1/2}=1/(1+m_c)^2 \approx 0.42$.
Given that $1/4 < v_{s=1/2} <1$, our theory indeed predicts
metamagnetism and multipolar phases for the FFHC, in agreement with
numerical observations \cite{Sudan2009}.

{\em Generalizations and Outlook:} The non-linear sigma model
formulation can be easily extended to higher-dimensional Lifshitz
points, and moreover, a rescaling argument similar to that in
\eqref{eq:4} continues to apply, implying that again asymptotically
exact solutions are possible.   This may provide a means to
understand other frustrated ferromagnets and ferrimagnets,
including possibly the kagom\'e lattice material volborthite
\cite{Ishikawa2015,Janson2015}, which shows signs of nematic-like
behavior below an unusually-wide $1/3$ magnetization plateau.

{\em Acknowledgements.} 
We would like to thank A. Furusaki for detailed discussions of magnon binding. 
We also thank A. Chubukov, T. Momoi, Z. Hiroi and M. Takigawa
for insightful discussions. Our work is supported by the NSF under
grant no. DMR-15-06119 (L.B.) and DMR-12-06774 (O.A.S.).  
This research benefitted from the facility of the KITP, supported by NSF grant PHY11-25915.

\bibliography{J1J2.bib}

\begin{widetext}

\renewcommand{\theequation}{S-\arabic{equation}}
  \setcounter{equation}{0}  
  \renewcommand{\thefigure}{S-\arabic{figure}}
   \setcounter{figure}{0}
\newpage
\begin{center}
\bf{Supplementary Information for
`` A Panoply of Orders from a Quantum Lifshitz Field Theory'' \\ Leon Balents and Oleg A. Starykh}
\end{center}

\section{Saddle point analysis}
The saddle point of Eq.2 with minimum action describes a cone
(umbrella) state:
\begin{equation}
  \label{eq:5A}
  \hat{m}_{\rm sp} = (\varphi\cos q x, \varphi\sin qx, \sqrt{1-\varphi^2}).
\end{equation}
The corresponding energy density is 
\begin{equation}
{\cal E}_{\rm cone} = - \delta \varphi^2 + \kappa \varphi^2 q^4 - \kappa v \varphi^4 q^4 - h\sqrt{1-\varphi^2}.
\end{equation}
Minimizing it over $q$ gives $q^2 = \delta/(2\kappa(1-v \varphi^2))$. Hence
\begin{equation}
{\cal E}_{\rm cone} = -\frac{\delta^2 \varphi^2}{4\kappa (1-v \varphi^2)} - h\sqrt{1-\varphi^2}.
\end{equation}
Energy density of the ferromagnetic phase, where $\varphi=0$, is just ${\cal E}_{\rm FM}=-h$.

At the first order transition, two conditions should be satisfied: (a) ${\cal E}_{\rm cone} = {\cal E}_{\rm FM}$, 
and (b) $\frac{\partial {\cal E}_{\rm cone}}{\partial \varphi^2} = 0$. 
The first one tells that 
\begin{equation}
b (1 - \sqrt{1-\varphi^2}) = \frac{\varphi^2}{2(1-v\varphi^2)}, 
\end{equation}
where $b = 2\kappa h/\delta^2 = h/h_0$, while the second leads to 
\begin{equation}
b =\frac{\sqrt{1-\varphi^2}}{(1-v\varphi^2)^{2}}.
\label{eq:b}
\end{equation}
Combining these two equations we find
\begin{equation}
  \label{eq:9A}
  \varphi_c^2 = \frac{2\sqrt{v}-1}{v}, h_c = \frac{\delta^2}{8 \kappa \sqrt{v}(1-\sqrt{v})}, ~{\text{and}}~ q_c^2 = \frac{\delta}{4\kappa(1-\sqrt{v})}.
\end{equation}
Observe that $h_c > h_0 = \delta^2/(2\kappa)$ for $v > 1/4$.

Condition (b) [Eq.\eqref{eq:b}] applies everywhere inside the cone phase and is used to find the magnetization $m=\sqrt{1-\varphi^2}$. Namely, 
it leads to $b^2 (1 - v \varphi^2)^4 = (v-1)/v + (1-v\varphi^2)/v$. In terms of $z = 1 - v \varphi^2$ this gives quartic equation
$b^2 z^4 - z/v + (1-v)/v = 0$, solution of which gives $z(b)$ for a given $v$.
We find that there is only one physical root in the entire interval $0 < v < 1$. Differentiating both sides of \eqref{eq:b} with respect to $b$
one finds relation between $m' = \partial m/\partial b$ and the magnetization, $m' = (1 - v + v m^2)^3/(1-v-3v m^2)$. Hence, near $m=0$
the slope of the magnetization (that is, spin susceptibility) is $(1-v)^2/h_0$. Near the saturation, $m=1$, the slope is $1/(h_0 (1 - 4 v))$.
In particular, at the critical $v=1/4$, separating the continuous from the discontinuous transitions, the slope diverges. For $v > 1/4$ magnetization
is continuous up to $m_c = \sqrt{1-\varphi_c^2} = (1-\sqrt{v})/\sqrt{v}$, which is reached at $h_c = h_0/(4 \sqrt{v}(1-\sqrt{v})) > h_0$. It jumps to
the saturation, $m_{\rm sat} = 1$, at $h=h_c +0$. This behavior is easily identifiable in Fig.1 of the main text.

To find the cone state energy at $h=h_0$, which is required in our analysis of the metamagnetic endpoint, we need to solve Eq.\eqref{eq:b} at $b=1$.
Assuming that corresponding order parameter $\varphi_0$ is proportional to $\varphi_c$ at the critical point, $\varphi_0^2 = a \varphi_c^2$, 
and expanding to second order in $\epsilon = v - 1/4 \ll 1$, we find $a=4/3$. It then easy to find that 
$\Delta {\cal E}(h_0) = {\cal E}_{FM}- {\cal E}_{\rm cone}|_{h_0} = 256\epsilon^3 \delta^2/(27 \kappa)$.

\section{Metamagnetic endpoint}
Magnon Hamiltonian 
\begin{equation}
H_{\rm fluc} = \sum_k 2A_k \bar{\eta}_k \eta_k + B_k (\eta_k \eta_{-k} + \bar{\eta}_k \bar{\eta}_{-k}) .
\label{eq:H_fluc2}
\end{equation}
leads to the zero-point energy $\delta {\cal E}_{\rm cone} = \int_{-\Lambda_0}^ {\Lambda_0} \frac{d k}{2\pi} \{ \sqrt{A_k^2 - B_k^2} - A_k\}$.

We find that
\begin{equation}
s A_k = \kappa k^4 + \frac{3 - 6 v^{1/2} + 6 v -4 v^{3/2}}{4v(1-v^{1/2})} k^2 + \frac{(3-14 v^{1/2} + 20 v - 8 v^{3/2})\delta^2}{32 \kappa (1-v^{1/2})^2 v}
\end{equation}
and
\begin{equation}
s B_k = \frac{\delta (2v^{1/2} -1)}{32 \kappa (1-v^{1/2})^2 v}\Big(8 \kappa (1-v^{1/2})(3-2v) k^2 + (3-8v^{1/2} + 4v)\delta\Big)
\end{equation}

In the limit $v=1/4 + \epsilon, \epsilon\to 0$ these simplify to 
\begin{eqnarray}
s A_k &=& \kappa k^4 + 2 \delta k^2 - 10 \delta k^2 \epsilon + (38 \delta k^2 + \frac{4\delta^2}{\kappa})\epsilon^2 + ...\nonumber\\
s B_k & = & 10 \delta k^2 \epsilon - (38 \delta k^2 + \frac{4\delta^2}{\kappa})\epsilon^2 + ...
\end{eqnarray}
Then, to the same $\epsilon^2$ accuracy, 
\begin{equation}
\{ \sqrt{A_k^2 - B_k^2} - A_k\} \to -\frac{50 \epsilon^2 \delta^2 k^2}{s(\kappa k^2 + 2\delta)} = -\frac{50 \epsilon^2 \delta^2}{s \kappa} + 
\frac{100 \epsilon^2 \delta^2}{s(\kappa k^2 + 2\delta)}
\end{equation}
As a result
\begin{equation}
\delta {\cal E}_{\rm cone} = -\frac{50 \epsilon^2 \delta^2 \Lambda_0}{s \kappa \pi} + \frac{25 \sqrt{2} \epsilon^2 \delta^{5/2}}{s \kappa^{3/2}},
\end{equation}
where in the second term the integration was extended to infinity due to its convergence.

Observe that the first term represents a regular correction $\delta {\cal E}_{\rm cone}^{\rm reg} = -\frac{50 \epsilon^2 \delta^2 \Lambda_0}{s \kappa \pi}$,
which scales with $\delta$ the same way as the saddle-point energy $\Delta {\cal E}(h_0) = \frac{256}{27} \frac{\epsilon^3 \delta^2}{\kappa}$.
It can be sought of as a renormalization of $\epsilon = v-1/4$ by quantum fluctuations.

The second term, on the other hand, is a singular $\delta {\cal E}_{\rm cone}^{\rm sing} = \frac{25 \sqrt{2} \epsilon^2 \delta^{5/2}}{s \kappa^{3/2}}$
correction which scales as fractional, ${5/2}$, power of $\delta$. It represents a physically distinct contribution to the cone state energy.

Thus, as described in the main text, $\Delta E = \Delta {\cal E}(h_0) - \delta {\cal E}_{\rm cone}^{\rm sing}$ turns to zero at 
$\delta_c = (\frac{256}{25 \cdot 27 \sqrt{2}})^2 \kappa s^2 \epsilon^2 \approx 0.07  \kappa s^2 \epsilon^2 \ll 1$.

\section{Two-particle Schrodinger equation in the continuum}
\label{sec:2particle}
Using parameterization eq.7 with $(\hat{e}_1, \hat{e}_2, \hat{e}_3) = (\hat{x}, \hat{y}, \hat{z})$, as appropriate for the fully
polarized vacuum state, we find that action eq.2 turns into
\begin{eqnarray}
S&=&\int dxd\tau ~\bar{\eta} \partial_\tau \eta + \cal{H}_\eta,\nonumber\\
{\cal H}_\eta &=& \frac{1}{s}\Big(h \bar{\eta} \eta - 2 \delta \bar{\eta}' \eta' + 2 \kappa \bar{\eta}'' \eta''\Big) + 
\frac{1}{2s^2} \Big( - \delta [\bar{\eta}^2 (\eta')^2 + (\bar{\eta}')^2 \eta^2] + 8 (1-v) \kappa (\bar{\eta}')^2 (\eta')^2 +\nonumber\\
&&4\kappa[ \eta \eta' \bar{\eta}' \bar{\eta}'' + \bar{\eta} \bar{\eta}' \eta' \eta''] - 2\kappa [ \bar{\eta} \bar{\eta}'' (\eta')^2 + \eta \eta'' (\bar{\eta}')^2] 
+ \kappa[ \bar{\eta}^2 (\eta'')^2 + \eta^2 (\bar{\eta}'')^2]\Big) ,
\end{eqnarray}
where prime stands for spatial derivative.

Fourier transforming gives 
\begin{eqnarray}
  \label{eq:6a}
  H  =  \sum_k \epsilon_k \overline{\eta}_k \eta_k +
  \frac{1}{2 L} \sum_{kpp'} V(k,p,p')\overline{\eta}_{k/2+p}
  \overline{\eta}_{k/2-p} \eta_{k/2-p'}\eta_{k/2+p'}, \nonumber
\end{eqnarray}
with $\epsilon_k = (h + 2\kappa k^4 - 2\delta k^2)/s$
and the symmetrized potential $V(k,p,q)$ is given by 
\begin{eqnarray}
V(k,p,q) &=& \frac{1}{s^2} \Big\{\frac{1}{2} \delta k^2 - \frac{1}{8} \kappa (1+4v) k^4 - \delta (p^2 + q^2) + \nonumber\\
&& + \kappa \Big(p^4 + q^4 + \frac{1}{2} (-3 + 4v) k^2 (p^2 + q^2) + 4(3 - 2v) p^2 q^2\Big)\Big\}
\end{eqnarray} 
Observe that as far as the spin $s$ dependence goes, $V(k,p,q)$ is actually $s$-independent, as it should be, because both $\delta$ and  $\kappa$
scale as $s^2$.
Observe also that the center of mass (CM) momentum of the pair $k$ couples to the relative momenta $p$ and $q$. In our case $k = 2q_0 = \sqrt{2\delta/\kappa}$.
Assuming for the moment that $p, q \ll q_0$, we can extract the constant (momentum-independent) part of the interaction 
\begin{equation} 
V(2q_0,0,0) = \frac{2\kappa (1- 4 v) q_0^4}{s^2} = \frac{(1-4v) \delta^2}{2 \kappa s^2} = \gamma_1.
\end{equation}
This is just $\gamma_1$, an attractive contact interaction (for $v > 1/4$) between magnons from the same valley.

The general form of a two-particle
state is $|\psi,k\rangle = \int \! \frac{dq}{2\pi}
\Psi(q;k)\overline{\eta}_{k/2+q}\overline{\eta}_{k/2-q} |0\rangle$,
where $|0\rangle$ is the boson vacuum, i.e. the ferromagnetic state,
$k$ is the (conserved) center of mass momentum, and the two-magnon
wavefunction obeys
\begin{eqnarray}
  \label{eq:7a}
 (\epsilon_{k/2+p} + \epsilon_{k/2-p} - E) \Psi(p;k) + \int \!
  \frac{dq}{2\pi} V(k,p,q) \Psi(q;k) =0.
\end{eqnarray}
The minimum energy state for $k = 2q_0$ is parameterized in terms of 
the bound state energy $\epsilon_b$ measured from the minimum of two-particle continuum at $k = 2q_0$ as $E = (2 h - 4 \kappa q_0^4)/s - \epsilon_b$. 
Then \eqref{eq:7} 
\begin{eqnarray}
&&\Big(s \epsilon_b + 16 \kappa q_0^2 p^2 + 4 \kappa p^4\Big)  \Psi(p;2q_0) = - s   \int \frac{dq}{2\pi} V(2q_0,p,q) \Psi(q;2q_0) \nonumber\\
&&=  - s   \int \frac{dq}{2\pi} \Big( \gamma_1 - \frac{8 \kappa q_0^2(1-v)}{s^2} (p^2 + q^2) + \frac{4 \kappa (3-2v)}{s^2} p^2 q^2 
+ \frac{\kappa}{s^2} (p^4 + q^4)\Big)  \Psi(q;2q_0)
\label{eq:8a}
\end{eqnarray}
In the very simple limit of $p, q \ll q_0$, which corresponds to the low-energy Hamiltonian eq.10, we are allowed to neglect all momentum
dependent terms in the integrand of the right-hand-side. Then
\begin{equation}
 \int \frac{dp}{2\pi} \Psi(p;2q_0) = \int \frac{dp}{2\pi} \frac{-s\gamma_1}{s\epsilon_b + 16 \kappa q_0^2 p^2} \int \frac{dq}{2\pi} \Psi(q;2q_0)
\end{equation}
which leads, for $\gamma_1 < 0$, to the bound state energy 
$\epsilon_{b0} = \kappa (1-4v)^2 q_0^6/(16 s^3) = (1-4v)^2 \delta^3/(128 s^3 \kappa^2) = \epsilon^2 \delta^3/(8 s^3 \kappa^2)$.
(Remember that $v = 1/4 + \epsilon$.)
This, of course, describes the bound state due to an attractive delta-function potential of strength $\gamma_1$.

The solution of the full equation \eqref{eq:8} is more complicated. We first turn it into 
a linear algebra problem $\psi_n = \sum_{m=0,2,4} A_{n,m} \psi_m$
for the {\em moments} $\psi_n = \int_{-\Lambda_0}^{\Lambda_0} \frac{dq}{2\pi} q^n \Psi(q;2q_0)$. The bound state energy $\epsilon_b$ is then found from 
${\text{det}}[ A - \hat{1}] = 0$. Note that the matrix elements of $A$ are formed by momentum integrals involving upto 8th power of 
momentum in the numerator and 4th power of momentum in the denominator of the integrand. This requires special care in treating divergent integrals.
The upper cut-off $\Lambda_0$ is such that $q_0 \ll \Lambda_0 \ll 1$. The first (left) inequality allows us to account for inter-valley scattering
with momentum transfer of the order $\pm 2q_0$, while the second (right) follows from the fact that the field theory is formulated in continuum
and is obtained by integrating out all lattice-scale fluctuations with wave vectors of order $1$ (the lattice spacing is set to 1 for convenience).
We proceed by carefully treating converging ($\Lambda_0$-independent) elements of $A$ and by separating singular (in $\epsilon_b \to 0$ limit)
elements there from order 1 contributions. The rest of matrix elements is organized in power series in $\Lambda_0 \ll 1$. 
Plugging these all back in ${\text{det}}[ A - \hat{1}] = 0$ we finally obtain
\begin{equation}
  \sqrt{\epsilon_b} \approx \sqrt{\epsilon_{b0}} \left[1 -
   \left(\frac{\delta}{\delta_{c2}}\right)^{1/2}\right]+ \mathcal{O}(\delta^{5/2}),
\end{equation}
Here $\delta_{c2} = 128 \kappa s^2 \epsilon^2/625$, and $\epsilon_{b0}$ is actually written in terms of the {\em renormalized} 
interaction parameter  $\tilde\epsilon = \epsilon + 25 \Lambda_0/(8 \pi s)$. The bound state disappears for $\delta > \delta_{c2}$.

\section{Parameters of the action}
\label{sec:action}

To find bare values of $\delta, \kappa$ and $\upsilon$ for the action $S$ (Eq.2 of the main text), 
we match, at $h=0$, single-particle dispersion as predicted by the field theory $\epsilon_k = (2\kappa k^4 - 2\delta k^2)/s$ (see eq.9 and discussion
below it) 
with that obtained directly from the lattice FFHC model, in the limit of small momentum $k$.
The latter one is given by $\omega_k = 2s( J_1 (\cos[k]-1) + J_2 (\cos[2k]-1)) \to 2s( (\frac{1}{2} - 2\beta) k^2 + (\frac{2\beta}{3} - \frac{1}{24})k^4)$,
where we set $J_1 = -1$ and $J_2 = \beta$. Hence $\delta = s^2 (4\beta -1)/2$ and $\kappa = s^2(\frac{2\beta}{3} - \frac{1}{24}) \to s^2/8$,
where we set $\beta=1/4$ in the expression for $\kappa$. Note that $\upsilon=0$ at this classical, $s^2$, level. 

{\em The saddle point analysis.} Using that $\varphi=0$ at $h=0$ and $\hat{m} = (\cos q x, \sin q x, 0)$ we find 
${\cal E}_{\rm cone}(q) = - \delta q^2 + \kappa q^4 + \upsilon (q^2)^2$.
Clearly, the optimal $q$ is $q_{\rm sp}^2 = \delta/(2(\kappa + \upsilon))$ and ${\cal E}_{\rm cone}(q_{\rm sp}) = - \delta^2/(4(\kappa + \upsilon))$. 
Hence, we can turn these relations around as
$\delta = - 2 {\cal E}_{\rm cone}(q_{\rm sp})/q_{\rm sp}^2$ and $\kappa + \upsilon = - {\cal E}_{\rm cone}(q_{\rm sp})/q_{\rm sp}^4$.
(Observe that ${\cal E}_{\rm fm} = 0$.)

Then, using the result $q = q_0 (1 + \frac{3}{4s})$ of the large-$s$ calculation described in the next Section~\ref{sec:largeS}, we obtain
\begin{equation}
\delta = \frac{2 s^2}{q_0^2 (1 + \frac{3}{4 s})^2} \frac{(J_1 + 4J_2)^2}{8 J_2} (1 + \frac{3}{2s}) = s^2 (2\beta - 1/2) + O(s^{0}),
\end{equation}
which means that $\delta$ is not changed by quantum fluctuations to our $1/s$ order.

Similarly, for $\kappa + \upsilon$ we get
\begin{equation}
\kappa + \upsilon = \frac{2 s^2}{q_0^4 (1 + \frac{3}{4 s})^4} \frac{(J_1 + 4J_2)^2}{8 J_2} (1 + \frac{3}{2s}) = s^2 (\frac{J_2}{2} - \frac{3 J_2}{4s}) + O(s^{0})),
\end{equation}
Assuming that $\kappa=s^2 J_2/2 = s^2/8$ does not renormalize (because, at $\delta =0$, it describes excitations of the state with no quantum fluctuations), we
see that $\upsilon = - 3 J_2 s/4 = -3 s/16$ and hence $v = - \upsilon/\kappa = 3/(2 s)$.

Thus,
\begin{equation}
\delta = \frac{s^2 (4\beta -1)}{2}, \kappa = \frac{s^2}{8}, \upsilon = -\frac{3 s}{16 }, v = \frac{3}{2s}.
\end{equation}

\subsection{Large-$s$ calculation of $v$}
\label{sec:largeS}

Our goal is to determine the quantum fluctuation term $\upsilon$ in the NLsM by comparing the wavevector of 
the exact ground state to that predicted by the NLsM. Of course we do not know the exact wavevector, but at least we can obtain the 
$1/s$ correction to the classical one. In this section, we discuss how to obtain such a $1/s$ correction to $q$.

The idea is to calculate the energy as a function of the ordering wavevector $q$, and minimize it.  A priori we expect that the energy density $e(q)$ has a series expansion,
\begin{equation}
  \label{eq:52}
  e(q) = s^2 e_0(q) + s e_1(q) + \cdots
\end{equation}
To minimize it, we require $e'(q)=0$.  We then let
\begin{equation}
  \label{eq:53}
  q = q_0 + \frac{1}{s} q_1 + \cdots
\end{equation}
We can collect the terms to the first two orders, which give the conditions
\begin{eqnarray}
  \label{eq:54}
  e'_0(q_0) & = & 0, \\
  e''_0(q_0) q_1 + e'_1(q_0) & = & 0.
\end{eqnarray}
The first condition just expresses that $q_0$ is the classical minimum, and the second determines the leading correction $q_1$, which is what we are after. 

The Hamiltonian
\begin{equation}
H = \sum_n J_1 \vec{S}_n \cdot \vec{S}_{n+1} + J_2 \vec{S}_n \cdot \vec{S}_{n+2}
\end{equation}
is studied by transforming into rotating grame
\begin{eqnarray}
\left( \begin{array}{c}
S^x_n \\
S^y_n \\
S^z_n \end{array} \right)
=
\left( \begin{array}{ccc}
\cos q n& -\sin q n & 0 \\
\sin q n & \cos q n & 0 \\
0 & 0 & 1  \end{array} \right) 
\left( \begin{array}{c}
S^1_n \\
S^2_n \\
S^3_n \end{array} \right)
\end{eqnarray}
and expanding spin in the local basis as
\begin{eqnarray}
\left( \begin{array}{c}
S^1_n \\
S^2_n \\
S^3_n \end{array} \right)
=
\left( \begin{array}{c}
S - a^\dagger_n a_n \\
\sqrt{\frac{S}{2}}(a^\dagger_n + a_n)  \\
i\sqrt{\frac{S}{2}}(a^\dagger_n - a_n) \end{array} \right).
\end{eqnarray}

This gives for the classical energy per spin
\begin{equation}
e_0(q) = J_1 \cos q + J_2 \cos 2q,
\end{equation}
minimization of which leads, of course, to $\cos q_0 = -J_1/(4J_2)$, so that $q_0 \approx \sqrt{8(\beta -1/4)}$.
The energy of the ferromagnetic state, $q=0$, is $e_{0;fm} = J_1 + J_2$.

Leading quantum fluctuations are described by 
\begin{equation}
H^{(2)} = S \sum_k \{ 2A_k a^\dagger_k a_k - B_k (a_k a_{-k} + a^\dagger_k a^\dagger_{-k})\}
\end{equation}
with 
\begin{eqnarray}
2 A_k &=& J_1 [(1+\cos q) \cos k - 2 \cos q] + J_2 [(1+\cos 2q) \cos 2k - 2 \cos 2q], \nonumber\\
2 B_k &=& J_1 (1-\cos q) \cos k + J_2 (1 - \cos 2q) \cos 2k .
\end{eqnarray}
Note that in the ferromagnetic state $q=0$ the anomalous part is absent, $B_k = 0$.

Diagonalizing $H^{(2)}$ we find the required zero-point motion energy
\begin{equation}
e_1(q) = \frac{1}{N}\sum_k (\Omega_k - A_k) = J_1 \cos q + J_2 \cos 2q + \frac{1}{N}\sum_k \Omega_k , 
\end{equation}
where $\Omega_k = \sqrt{A_k^2 - B_k^2}$. 
As a result, we need to calculate $e_1'(q_0) = \frac{1}{N}\sum_k \frac{\partial \Omega_k}{\partial q} |_{J_1 = -4 J_2 \cos q_0}$.
That is, take derivative of $\Omega_k$ over $q$, and then make the substitution
$J_1 = -4 J_2 \cos q_0$. The obtained result can then be expanded 
in powers of $q_0 \ll 1$ (which is well
justified near the Lifshitz point) to the 3rd order and integrated over $k$. 
In this way we find $e_1'(q_0) = - 3 J_2 q_0^3$ and $q_1 = 3q_0/4$, so that $q = q_0 (1 + \frac{3}{4s})$.

Observe that by virtue of the relation $\delta \sim q_0^2$, this result implies the scaling $e_1 \sim q_0^4 \sim \delta^2$.
However, calculation of the higher order in $\delta$ terms, of the type $\delta^n$ with $n \geq 3$, results in infrared divergent integrals.
These divergencies
imply that the whole perturbation series needs to be re-summed in order to obtain a finite $\delta^{5/2}$
contribution found in the main text. Luckily, large-$s$ calculation of the interaction parameter $v$ does not require these
higher powers of $\delta$.

To summarize, large-$s$ calculation predicts $q = q_0 (1 + \frac{3}{4s})$ and 
cone state energy, relative to that of the ferromagnetic state, $e(q) = s^2 (e_0 - e_{0;fm}) + s e_1 = - s^2 \frac{(J_1 + 4J_2)^2}{8 J_2} (1 + \frac{3}{2s})$.

\end{widetext}
\end{document}